\documentclass[aps,prb,reprint,superscriptaddress]{revtex4-1}

\usepackage{graphicx}
\def \bise {Bi$_2$Se$_3$}
\def \bite {Bi$_2$Te$_3$}
\def \bitese {Bi$_2$(Te$_{1-x}$Se$_x$)$_3$}
\def \cm {cm$^{-1}$}

\def \etal{{\it et al.}}
\def \eg{E_g^{\rm direct}}
\renewcommand{\epsilon}{\varepsilon}
\begin{document}

%\preprint{}


\title{
	\boldmath
	Interband absorption edge in topological insulators 
 \bitese
\unboldmath}


\author{A. Dubroka}
\email[]{dubroka@physics.muni.cz}
%\thanks{}

\affiliation{Department of Condensed Matter Physics and Central European Institute of Technology, Masaryk University, Kotl\'a\v{r}sk\'a 2, 611 37 Brno, Czech Republic}


\author{O. Caha}
\affiliation{Department of Condensed Matter Physics and Central European Institute of Technology, Masaryk University, Kotl\'a\v{r}sk\'a 2, 611 37 Brno, Czech Republic}

\author{M. Hron\v{c}ek}
\affiliation{Department of Condensed Matter Physics and Central European Institute of Technology, Masaryk University, Kotl\'a\v{r}sk\'a 2, 611 37 Brno, Czech Republic}

\author{P. Fri\v{s}}
\affiliation{Department of Condensed Matter Physics and Central European Institute of Technology, Masaryk University, Kotl\'a\v{r}sk\'a 2, 611 37 Brno, Czech Republic}

\author{M. Orlita }
\affiliation{Laboratoire National des Champs Magn´etiques Intenses, CNRS-UGA-UPS-INSA-EMFL, 25, avenue des Martyrs,
38042 Grenoble, France}
\affiliation{Institute of Physics, Charles University in Prague, CZ-121 16 Prague, Czech Republic}

\author{V. Hol\'{y}}
\affiliation{Department of Condensed Matter Physics and Central European Institute of Technology, Masaryk University, Kotl\'a\v{r}sk\'a 2, 611 37 Brno, Czech Republic}
\affiliation{Department of Condensed Matter Physics, Charles University, Prague, Czech Republic}

\author{H. Steiner}
\affiliation{Institut f\"{ü}r Halbleiter- und Festk\"{ö}rperphysik, Johannes Kepler Universit\"{ä}t, Altenbergerstrasse 69, 4040 Linz, Austria}

\author{G. Bauer}
\affiliation{Institut f\"{ü}r Halbleiter- und Festk\"{ö}rperphysik, Johannes Kepler Universit\"{ä}t, Altenbergerstrasse 69, 4040 Linz, Austria}

\author{G. Springholz}
\affiliation{Institut f\"{ü}r Halbleiter- und Festk\"{ö}rperphysik, Johannes Kepler Universit\"{ä}t, Altenbergerstrasse 69, 4040 Linz, Austria}

\author{J. Huml\'\i\v{c}ek}
\affiliation{Department of Condensed Matter Physics and Central European Institute of Technology, Masaryk University, Kotl\'a\v{r}sk\'a 2, 611 37 Brno, Czech Republic}

\date{\today}

\begin{abstract}
We have investigated 
the optical properties of thin films of topological insulators Bi$_{2}$Te$_{3}$, Bi$_{2}$Se$_{3}$ and their alloys  \bitese\  on BaF$_{2}$ substrates by a combination of infrared ellipsometry and reflectivity in the energy range from 0.06 to 6.5 eV. 
For the onset of interband absorption in \bise, after the correction for the Burstein-Moss effect, we find the value of direct bandgap  of $215\pm10$~meV at 10~K. Our data supports the picture that \bise\ has a direct band gap located at the $\Gamma$ point in the Brillouin zone and that the valence band reaches up to the Dirac point and has the shape of a downward oriented paraboloid, i.e. without a camel-back structure. In \bite, the onset of strong direct interband absorption at 10~K is at a similar energy of about 200 meV, with a weaker additional feature at about 170~meV. Our data support the recent GW band structure calculations suggesting that the direct interband transition does not occur at the $\Gamma$ point but near the Z--F line of the Brillouin zone. In the \bitese\ alloy, the energy of the onset of direct interband transitions exhibits a maximum near $x=0.3$ (i.e. the composition of Bi$_2$Te$_2$Se), suggesting that the crossover of the direct interband transitions between the two points in the Brillouin zone occurs close to this composition. 
\end{abstract}

\pacs{75.47.Lx, 75.50.Cc, 71.30.+h}
%71.30.+h 	Metal-insulator transitions and other electronic transitions
%75.47.Lx 	Magnetic oxides
%75.50.Cc 	Other ferromagnetic metals and alloys

%\keywords{}

\maketitle

\section{Introduction}
Topological insulators belong to a class of materials that are insulating in the bulk, however on the surface they exhibit topological spin polarized conducting states\cite{Hsieh08,Taskin09,Xia2009,Nishide2010}. Since the spin polarization is potentially usable in spintronic devices, topological insulators have attracted a lot of attention in the last decade. \bitese\ compounds, originally employed for thermoelectric devices due to their large Seebeck constant\cite{Gordiakova1958}, belong to this category, thanks to the large spin orbit coupling due to heavy bismuth atoms. The linearly dispersing surface states have been observed by angle resolved photoemission (ARPES)\cite{Xia2009,Chen2009,Chen2010,Miyamoto2012, Neupane2012}. Samples are usually
of $n$-type with the Fermi level located in the conduction band because of the antisite doping related to Se vacancies~\cite{Horak1990}.

The infrared and optical response of \bitese\ compounds have been intensively studied in order to characterize their vibrational properties\cite{Richter1977}, interband absorption\cite{Greenaway1965,Kohler74Burnstein,LaForge2010,Chapler2014,Post2015PRB, Akrap2012} and free charge carrier properties\cite{Stordeur92,Post2013,DiPietro2012, LaForge2010}. Even the surface states\cite{Aguilar2012,Reijnders2014,Post2015PRL,Shao2017} and strong Faraday rotation\cite{Ohnoutek2016}  have been observed by infrared spectroscopy.  Despite the large experimental effort, several questions remain open. 
For example controversial results on the values of the band gaps determined by optical spectroscopy and ARPES were reported\cite{Post2013,Chapler2014}, raising questions about the actual position of the direct interband transition onset in the Brillouin zone. Here we present infrared spectroscopy study of \bitese\ thin films that focuses on the absorption edge of interband transitions. We discuss and account for the Burstein-Moss (BM) effect in order to obtain bandgap energies. We compare our results with ARPES data and band structure calculations and discuss the location of the onset of direct interband transitions in the Brillouin zone.

The paper is organized as follows: in Sec.~\ref{Techniques} we describe  the sample growth, X-ray diffraction results and details of the analysis of the optical data. The core of the paper is presented in Sec.~\ref{Discussion}, where we discuss the results concerning the onset of interband absorption in \bise\ (Sec.~\ref{BiSeText}), \bite\ (Sec.~\ref{BiTeText}) and finally in \bitese\ alloys (Sec.~\ref{BiTeSeText}).

\section{Experimental techniques}
\label{Techniques}
\subsection{Sample preparation and X-ray diffraction}

\bitese\ epilayers were grown by molecular beam epitaxy in a Riber 1000 system under ultrahigh vacuum conditions at a background pressure smaller than $5\times10^{-10}$\,mbar.
The molecular beams were generated using compound bismuth telluride or bismuth selenide effusion cells (nominal
composition of \bite\ and \bise, respectively) operated at around 400--500$^\circ$C, and separate
tellurium or selenium cells operated at around 200--300$^\circ$C for stoichiometry control.
The ternary alloys were grown combining \bise{} and Te effusion cells (low Se content) or \bite\ and Se effusion cells (high Se content), respectively.
The particular flux rates of the compounds were around 1\,\AA/s for compound cells and 2--3\,\AA/s for tellurium or selenium, which were calibrated by a quartz crystal microbalance.
The layers were deposited on 1~mm thick cleaved BaF$_2$ (111) substrates  at temperature between 300--400\,$^\circ$C as measured with an infrared optical pyrometer.
The surface structure of the films was monitored by in situ reflection high-energy electron diffraction evidencing 2D growth for all samples under the given growth conditions. The \bite\ and \bise\ thin films from this growth series were recently characterised by X-ray and Raman spectroscopy~\cite{Caha2013,Humlicek2013}.

\begin{figure}
	\hspace*{-1cm}
	\includegraphics[width=8cm,angle =-90 ]{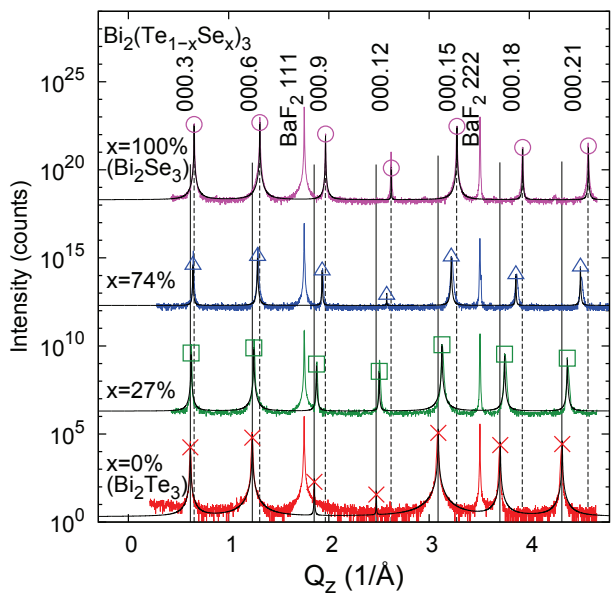}%
	\caption{\label{xrd} 
		Symmetric X-ray diffraction scans of our \bitese\ samples ordered from bottom to top with increasing selenium content recorded with Cu K$\alpha_1$ radiation. The spectra are shifted for clarity with respect to each other. The corresponding diffraction peaks shift to the higher  $Q_z$ values with increasing selenium content $x$ and decreasing $c$ lattice constant. 	Solid and dashed lines show positions of the \bite\ and \bise\ diffraction peaks, respectively. The symbols denote calculated structure factors of the \bitese{} lattice assuming a random occupation of Te and Se atoms. Thin black lines represent simulated diffraction profiles corresponding to the calculated structure factors.}
\end{figure}

The crystalline quality and composition of our thin films were determined using X-ray diffraction. The symmetric scans along $(000l)$ reciprocal space direction are shown in Fig.~\ref{xrd} with respect to the scattering vector $Q_z=4\pi /\lambda \sin\theta$, where $\lambda=1.54056\,$\AA{} is the  radiation wavelength and $\theta$ is the Bragg angle. The diffraction peaks of the \bitese\ alloy  shift towards higher $Q_z$ values with increasing selenium content and decreasing lattice parameter. The peaks are narrow in the whole composition range indicating that no phase separation occurs in the studied samples. Since the structure factor of the 000.9 and 000.12 diffractions in \bite\ is very small, the corresponding diffraction peaks are very weak. The chemical composition $x$ of the \bitese\ alloy was determined from the 
$c$ lattice parameter (see Tab.~\ref{Table}) assuming a linear concentration dependence of the lattice parameter (Vegard's law).
The intensities of the diffraction peaks in Fig.~\ref{xrd} correspond well to  calculated structure factors of random \bitese{} alloys, i.e. with the same probability of Se and Te atoms occupying all of the anionic positions within the crystalline structure. The solid lines in Fig.~\ref{xrd} show simulations of the diffraction spectra with the corresponding structure factors and peak shape given by Voigt profile with constant width of 0.003\,\AA$^{-1}$ evidencing an excellent agreement with the measured data. The width corresponds to the coherently diffracting domains mean size of 100\,nm.
Sharp diffraction peaks indicate high crystalline quality of the layers and homogeneous chemical composition in the layer.
\begin{table}[b]
	\caption{\label{Table}%
		Composition, $x$, lattice parameter along the $c$-axis, $c$, thickness, $d$, square  of plasma frequency, $\omega_{\rm pl}^2$, and phonon frequency $\omega_{\rm \alpha}$ and $\omega_{\rm \beta}$ of our \bitese\ thin films determined from 300~K data. The uncertainty of these values are typically 2\% for $x$, 0.05~\AA\ for $c$, 2\% for $d$,  5\% for $\omega_{\rm pl}^2$ and 2~${\rm cm}^{-1}$ for $\omega_{\rm \alpha}$	and $\omega_{\rm \beta}$ .
	}
	\begin{ruledtabular}
		\begin{tabular}{cccccc}
			$x$ & $c$   & $d$  & $\omega_{\rm pl}^2$  
			& $\omega_{\rm \alpha}$	& $\omega_{\rm \beta}$  \\		
			{}[\%] 	& 	[\AA] &	 [nm]	& [$10^6$ ${\rm cm}^{-2}$] 
			& [${\rm cm}^{-1}$] & [${\rm cm}^{-1}$]\\
			\colrule
			0  & 30.54  & 267 & 5.2 & 52 &  95 \\
			27 & 30.06 & 519 & 3.1 & 63 &115 \\
			74 & 29.23 & 207 & 1.1 & 66 &-- \\
			100 & 28.75 & 458 & 3.0 & 68 & 129\\
		\end{tabular}
	\end{ruledtabular}
\end{table} 

\subsection{The optical spectroscopy and analysis of optical data}
\begin{figure}
	\hspace*{-1.5cm}
	\includegraphics[width=11.5cm]{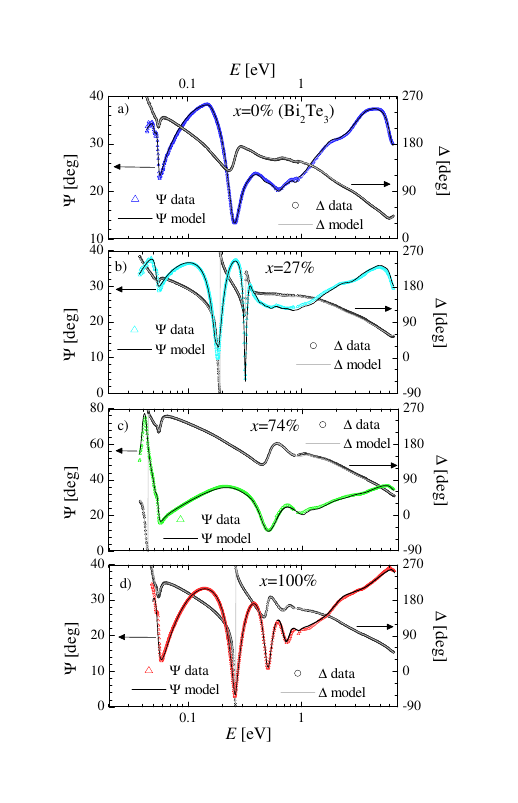}%
	\vspace*{-1.5cm}
	\caption{\label{elli}Measured ellipsometric angles $\Psi$ and $\Delta$ at 300~K (symbols) and model spectra  (solid lines) of our \bitese\ samples with $x=0\%$ (a), 27\% (b), 74\% (c) and 100\% (d) as a function of photon energy $E$. The angle of incidence was 70$^\circ$.}
\end{figure}

\begin{figure}
	\hspace*{-0.8cm}
	\includegraphics[width=10cm]{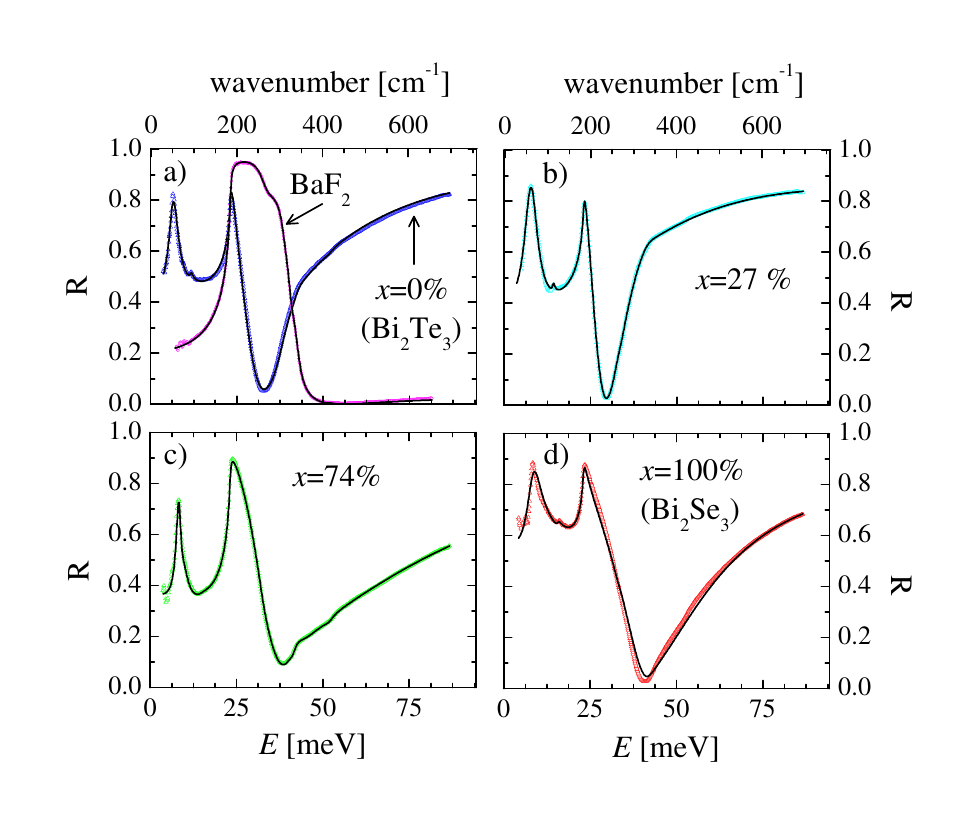}%
	\vspace*{-0.5cm}
	\caption{\label{R} Far-infrared reflectivity of our \bitese\ thin films on BaF$_2$ substrate at 300~K (symbols) together with model spectra (solid lines). Shown are results for samples with $x=0\%$ (a), 27\% (b), 74\% (c) and 100\% (d). In addition, the data (magenta circles) and the model spectrum (solid line) of the BaF$_2$ substrate are shown in panel (a). 
				}
\end{figure}

The optical properties were probed from the mid-infrared to ultraviolet  range  using a combination of two commercial ellipsometers Woollam IR-VASE  (60 -- 700~meV) and Woollam VASE (0.6--6.5~eV) equipped with variable  retarders. In the ellipsometric experiments, the two ellipsometric angles $\Psi$, $\Delta$  and depolarization are recorded\cite{Handbook}. The angles are linked to the Fresnel reflection coefficients for the p- and s-polarized wave, $r_{\rm p}$ and $r_{\rm s}$, respectively, as $\tan\Psi {\rm e}^{{\rm i}\Delta}=r_{\rm p}/r_{\rm s}$. From the two angles, using a proper model for the layered structure, the real and imaginary part of the dielectric function can be determined without the need of Kramers-Kronig relations. The room temperature spectra of $\Psi$ and $\Delta$  with respect to the photon energy $E$ are shown in Fig.~\ref{elli} (symbols) together with those of a model (solid lines) described below. 
The ellipsometric data were complemented in the far-infrared range (4--85~meV, 30--680~\cm) with near normal incidence reflectance (see Fig.~\ref{R}) measured with a Bruker Vertex 80v Fourier transform spectrometer. A gold mirror was used as a reference. The low-temperature data were acquired in the mid-infrared range with the Woollam IR-VASE  ellipsometer using a closed He-cycle cryostat. The cryostat was equipped with an ultra-low vibration interface in order to decouple vibrations of the Gifford-McMahon refrigerator from the ellipsometer. The pressure in the cryostat chamber was 1$\times10^{-7}$~mbar at 300~K.

The data were analysed with a model of coherent interferences in a thin film on a substrate. The dielectric function of BaF$_2$ was obtained from measurements on  a bare substrate, e.g. for far-infrared data see Fig.~\ref{R}(a). In the model we have taken into account also the incoherent reflections from the back side of the BaF$_2$ substrate that is cleaved and thus reflecting. The surface roughness was taken into account using an effective medium approximation (Bruggeman model)~\cite{Handbook}. The thickness of the surface roughness layer obtained from the fitting was between 5--11~nm, in good agreement with r.m.s. roughness value between 4--9~nm obtained  by atomic force microscopy. The surface roughness correction is increasingly important with decreasing wavelength of light $\lambda$, e.g., it significantly influences the dielectric function above about 1~eV ($\lambda\approx1\mu$m), however it has negligible impact at lower energies. As the first step, we analysed the response of the thin films in the whole measured frequency range.  We have modelled the dielectric function of the layer as the sum 
\begin{equation}
\hat{\epsilon}(\omega)=1-\frac{\omega_{\rm pl}^2}{\omega(\omega+{\rm i}\gamma)} 
+ \sum_{j}\frac{\omega_{{\rm pl},j}^2}{\omega^2-\omega^2_j-{\rm i}\gamma_j}
+\sum_{j}\hat{G}_j(\omega)
\end{equation}
where the angular frequency, $\omega$, relates to the photon energy, $E$, as 
$E=\hbar\omega$. The model consists of the Drude term for itinerant electrons (second term on the right hand side), Lorentz oscillators for transverse optical phonons (third term on the right hand side) and a set of Gaussian oscillators $\hat{G}_j(\omega)$ for interband transitions\cite{Handbook}.   The fit of model spectra to the data was performed using the Woollam WVASE software. The obtained best fit model spectra, that are displayed in Figs.~\ref{elli} and ~\ref{R} as solid lines, are in a very good agreement with the data (symbols).  The main features in the raw spectra in the far-infrared frequency range are (see Fig.~\ref{R}):
the strong (so-called $\alpha$) phonon between 52--68~\cm, the much weaker (so-called $\beta$) phonon between 95--130~\cm, and the BaF$_2$ substrate phonon at 186~\cm. The frequencies of the $\alpha$ and $\beta$ phonons (see Tab.~\ref{Table}) compare well with reported values~\cite{Richter1977,Post2013,Chapler2014}.
Between about 0.1--0.8 eV (see Fig.~\ref{elli}),  the films are transparent and the raw ellipsometric data  exhibit fringes due to interference in the layer. Above this range, the thin films are opaque and the features are solely due to the electronic interband transitions.

First we aimed at obtaining the plasma frequency of the itinerant charge carriers, $\omega_{\rm pl}$, and thickness of the layer, $d$. The results are summarized in Tab.~\ref{Table}. The thickness values were consequently used in the point by point retrieval of the dielectric function in the frequency range of ellipsometric data. We have assumed that the response of the layer is isotropic. This is certainly an approximation for the anisotropic response of \bitese\ caused by the rhombohedral crystal structure. In general, in the oblique reflectance geometry of an ellipsometric measurement, the out-of-plane component of the dielectric function is probed as well, although the main contribution to the data usually corresponds to the in-plane component. However, if the value of index of refraction is high, the wave refracts close to the sample normal and the anisotropic corrections are quite small unless the out-of-plane dielectric function has strong resonances in the loss function\cite{Bernhard2004,Foltin1992}. These requirements are fulfilled between 0.1--0.7~eV where the index of refraction is high (about 6--7). 
%Using the reported $c$-axis dielectric function~\cite{Richter1977}, we have checked that the anisotropy corrections in our case are indeed very small in the relevant energy range.

\section{Discussion}
\label{Discussion}
\subsection{Overview of the absorption spectra}

\begin{figure}
\hspace*{-0.5cm}
\includegraphics[width=10cm]{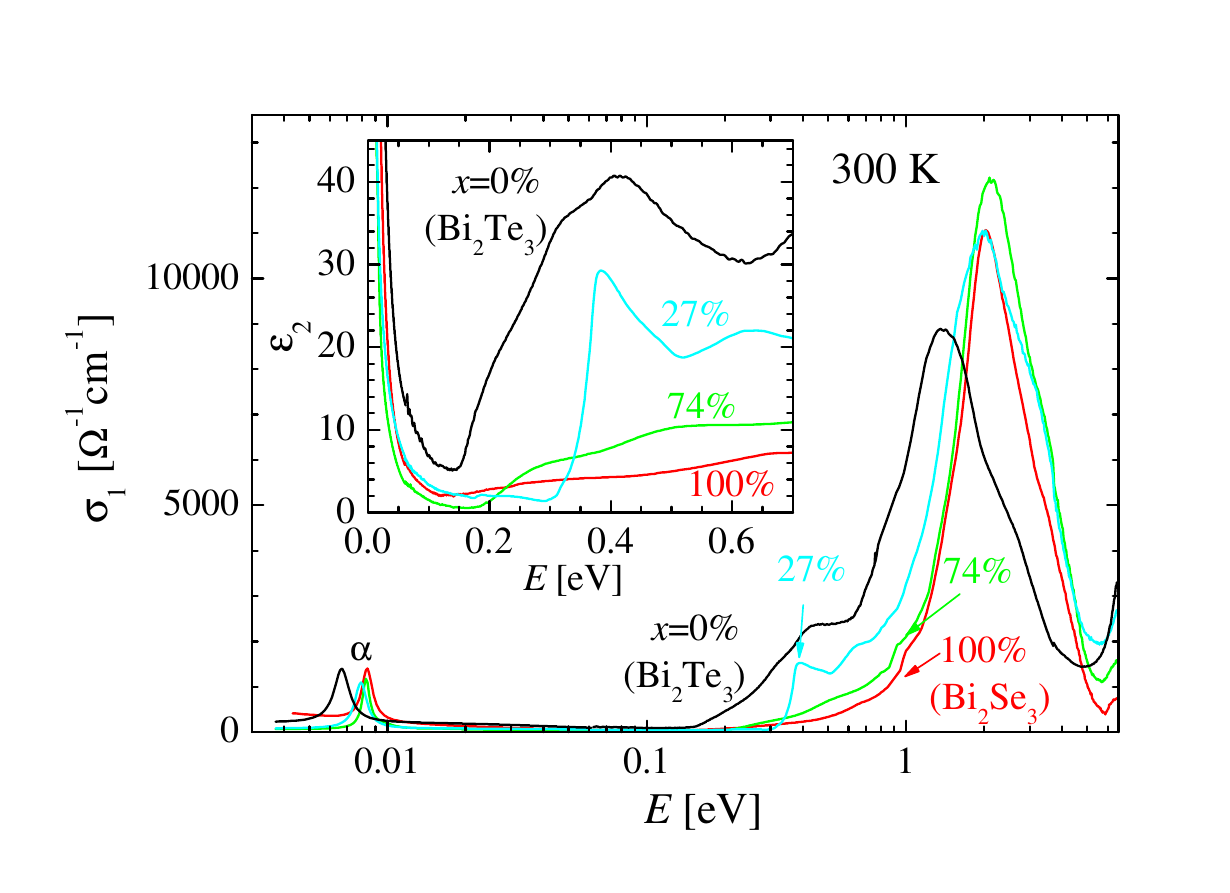}%
\caption{\label{s1}The real part of the optical conductivity at 300 K of the \bitese\ samples in the whole measured frequency range. The inset displays the region of the onset of interband absorption in terms of $\epsilon_2$.}
\end{figure}

Figure~\ref{s1} shows the spectra of our \bitese\ thin films at room temperature in the whole measured spectral range in terms of the real part of the optical conductivity $\sigma_1(\omega)=-{\rm i}\omega\epsilon_0\epsilon_2(\omega)$, where $\epsilon_0$ is the permittivity of vacuum and $\epsilon_2(\omega)$ is the imaginary part of the dielectric function. The prominent absorption structures are due to the $\alpha$ phonon between 6--9 meV, a weak free charge carrier response and the interband transitions setting in above $\approx0.1$~eV. In the following, we focus on the region of interband absorption edge between 0.1--0.7 eV, displayed in the inset of Fig.~\ref{s1} in terms of $\epsilon_2$. Above the bandgap, the maximum values of $\epsilon_2$ range from about 5 for the least absorbing Bi$_2$Se$_3$ sample up to about 40 for the most absorbing Bi$_2$Te$_3$ sample. These high values are typical for direct interband transitions. Indirect transitions have values of $\epsilon_2$ typically several orders of magnitude smaller~\cite{Cardona}. Therefore all the interband transitions discussed in this paper are direct and we designate their onset energy as $\eg$.

\begin{figure}
	\hspace*{-0.5cm}
	\includegraphics[width=9cm]{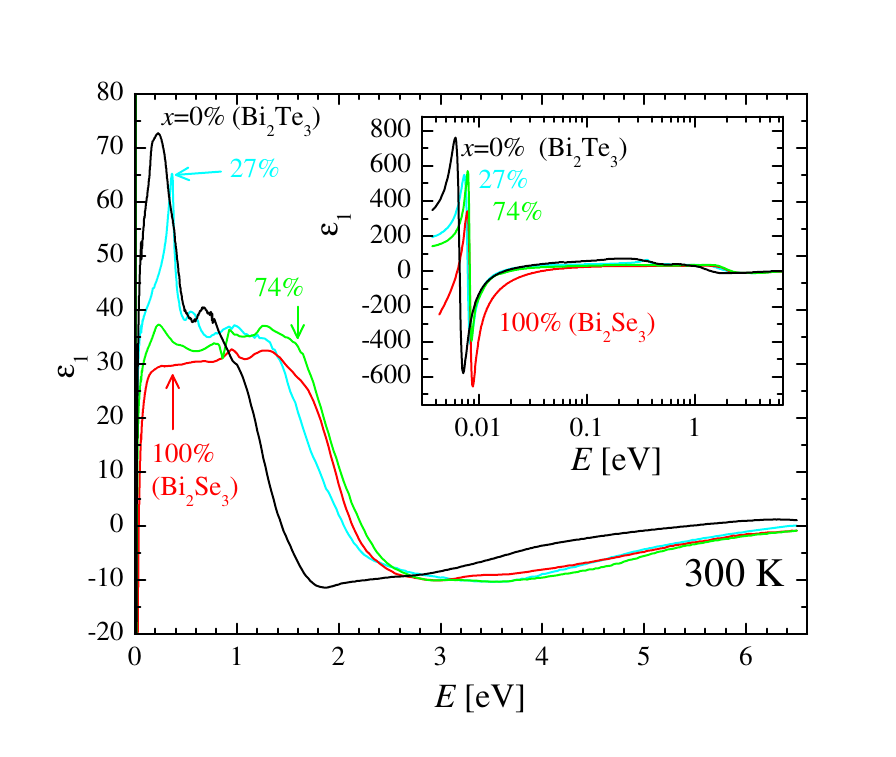}%
	\caption{\label{e1}The real part of the dielectric function, $\epsilon_1$,  at 300~K of the \bitese\ samples above the phonon energy range. The inset displays the data on semi-logarithmic scale including the large polarizability due to the $\alpha$-phonon.}
\end{figure}

The real part of the dielectric function, $\epsilon_1$, shown in Fig.~\ref{e1}, exhibits large values below the interband transitions ($\sim0.2$~eV) that range from 30 for \bise\ up  to more than 70 for \bite. 
The zero-crossing of $\epsilon_1$  above 5~eV indicates the renormalized plasma resonance of all valence electrons. The electronic polarizability is essentially due to the interband transitions below $\sim 4$~eV; the contribution of the interband transitions above the measured frequency range to the low energy value of $\epsilon_1$ is relatively small (about 1). At the lowest measured energy ($\sim 5$~meV, see the inset of Fig.~\ref{e1}), $\epsilon_1$ reaches several hundreds due to the low-lying $\alpha$-phonon.

\subsection{The absorption edge in \boldmath Bi$_2$Se$_3$ \unboldmath}
\label{BiSeText}

\begin{figure}
	\hspace*{-0.6cm}
	\includegraphics[width=9.8cm]{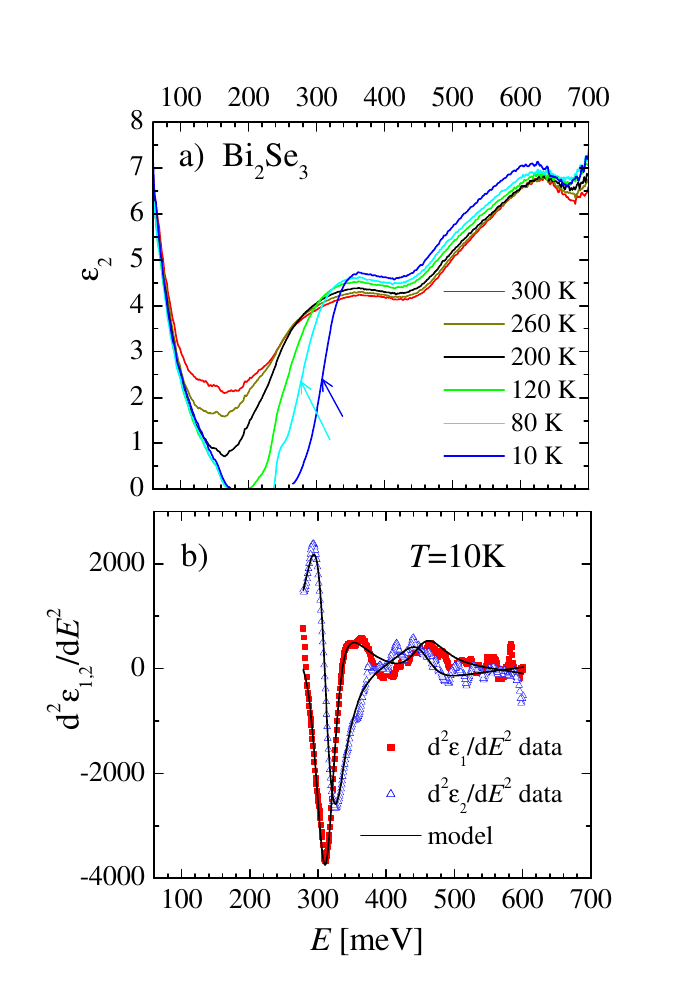}%
	\caption{\label{BiSe} a) Temperature dependence of the imaginary part of the dielectric function, $\epsilon_2$, of \bise. The arrows mark the energy of the lowest critical point at 10~K and 80~K, respectively. b) 
	Second derivative of the real (red squares) and imaginary part (blue triangles) of the dielectric function at 10~K together with the spectrum of the critical point model (solid lines) specified in the text.}
\end{figure}

Figure~\ref{BiSe}(a) shows the temperature dependence of $\epsilon_2$ of the \bise\ thin film in mid-infrared range. The onset of interband transitions with lowering temperature sharpens and shifts to higher energies similarly to previously published results~\cite{Post2013}. The weak feature near 250~meV, which is most visible in the 80~K spectrum, is an artefact due to the sharp interference structure at this energy [see Fig.~\ref{elli}(d)]. Below the onset of interband transitions at low temperature, $\epsilon_2$ is zero within the uncertainty of the analysis. This is consistent with the observed transmission values up to $14\%$ measured on 225~$\mu$m thick single crystal\cite{Ohnoutek2016}, which is possible only if $\epsilon_2\ll 0.1$. 

Due to a significant concentration of free electrons, the Fermi level, $E_{\rm F}$, lies in the conduction band. Consequently the minimal energy necessary to excite the electrons from the valence to conduction band, $E_{\rm min}$, is larger than the band gap $\eg$ because of the Pauli blocking of the states below $E_{\rm F}$. The latter effect is known as the Burstein-Moss (BM) effect or shift~\cite{Burstein54,Moss54}.
In such a case, a step in the absorption is expected. Indeed, it can be seen in Fig.~\ref{BiSe}(a) that  at 10~K, $\epsilon_2(E)$  has a step like shape reaching the value of about 5 in a rather narrow interval 290--340 meV. The representative energy of such an onset can be estimated as the middle point of the step, $\approx310$~meV, as shown by the arrow. In a more sophisticated way the onset can be determined with the help of the critical point model fitted to the second derivative of the dielectric function~\cite{Cardona1969, Humlicek2002}. The contribution of a parabolic critical point (CP) located at an energy $E_{\rm CP}$ to the $j$-th derivative of the dielectric function is
\begin{equation}
\label{CP}
\frac{{\rm d}^j\hat{\epsilon}(E)}{{\rm d}E^j}=Ae^{{\rm i}\phi}(E-E_{\rm CP}+{\rm i}\Gamma)^{-n-j}\;,
\end{equation}
where $A$ and $\phi$ are amplitude and phase factor and $n$ has the values -1/2, 0 and 1/2 for three-, two- and one-dimensional CP, respectively.
We have modelled the second derivative ($j=2$) of the dielectric function  shown in Fig.~\ref{BiSe}(b) with a sum of two CP functions of Eq.~(\ref{CP}). The data below 280~meV were omitted because of spurious structures due to the interference mentioned above. The most pronounced structure centred at $312$~meV was modelled with a two dimensional CP ($n=0$) that corresponds to the step in $\epsilon_2$~\cite{Handbook}. Its centre energy is very close to the middle point estimated above. 
The weak structure near 450~meV was modelled with a three dimensional CP.

Assuming for simplicity that the conduction and valence bands are parabolic, $E_{\rm min}$  and $\eg$ are related as
\begin{equation}
\label{BM}
E_{\rm min} = \eg + \frac{\hbar^2 k_{{\rm F},a}^2}{2} \left(\frac{1}{m_{\rm e}^a}+ \frac{1}{m_{\rm h}^a}\right)\;,
\end{equation}
where  $k_{{\rm F},a}$ is the in-plane Fermi wavevector and $m_{\rm e}^a$ and $m_{\rm h}^a$ are the in-plane effective masses of the conduction and valence band, respectively~\cite{Kohler74Burnstein}. With the help of the expression for the Fermi level with respect to the bottom of the conduction band, $E_{\rm F}=\hbar^2 k_{{\rm F},a}^2/(2m_{\rm e}^a)$,  Eq.~\ref{BM} can be rewritten as
\begin{equation}
E_{\rm min} = \eg + E_{\rm F}\left(1+\frac{m_{\rm e}^a}{m_{\rm h}^a}\right)\;.
\end{equation}

Since \bise\ is anisotropic, the equi-energy contour at the Fermi level has an ellipsoidal shape in the $k$-space. For the case of biaxial anisotropy, the Fermi level can be calculated from the concentration of charge carriers $n$ as
\begin{equation}
E_{\rm F} = 
\frac{\hbar^2}{2({m_{\rm e}^am_{\rm e}^b m_{\rm e}^c})^{1/3}}
	\left(3\pi^2n\right)^{2/3}\;,
\end{equation}
where $m_{\rm e}^a,m_{\rm e}^b$ 
and $m_{\rm e}^c$ are the three effective masses. 
In case of the uniaxial anisotropy of \bise, $m_{\rm e}^a=m_{\rm e}^b$.
We have determined the concentration of charge carriers from their plasma frequency $\omega_{\rm pl}$ (see Tab.~\ref{Table}) as
$
n = \omega_{\rm pl}^2m_{\rm e}^a\epsilon_0/e^2\;,
$
where $e$ is the elemental charge. For the evaluation of the BM shift, the following values of effective masses were used: $m_{\rm e}^a=0.14~m$~\cite{Orlita2015}, and $m_{\rm h}^a=0.24~m$\cite{Piot16}, where $m$ is the free electron mass. The $c$-axis effective mass can be calculated from the effective mass anisotropy  $m_{\rm e}^c/m_{\rm e}^a=1.6$ obtained for a Fermi energy $\approx50$~meV\cite{Kohler73BiSe} above the conduction band minimum, which is relevant for our sample. With these values, the BM shift amounts to 98 meV and the resulting band gap value is $\eg=215\pm10$~meV at 10~K. The uncertainty is estimated from that of the effective masses.  The calculation of the BM shift is based on the parabolic profile of the conduction and valence bands. The latter assumption is reasonably fulfilled for our relatively weakly doped sample with $n=4.8\times10^{18}\ {\rm cm}^{-3}$ which is below the threshold of $\approx10^{19}\ {\rm cm}^{-3}$, above which a significant nonparabolicity of the conduction band has been observed~\cite{Kohler73BiSe}. Similarly, the valence band was reported~\cite{Piot16} to have parabolic shape for the Fermi wavevector corresponding to our sample ($k_{{\rm F},a}\approx0.04$~\AA$^{-1}$).

%Note that the anisotropy $m_{\rm e}^c/m_{\rm e}^a$ depends on the doping level and values up to 4 were reported for some high dopings\cite{Stordeur92}. With the latter value of anisotropy, the BM shift would be 80~meV and the  band gap value $\eg=210$~meV.

In the following we compare the obtained value of $\eg$ with literature values for the band gap determined by optical spectroscopy and corrected for the BM shift. In their early study, K\"{o}hler and Hartman\cite{Kohler74Burnstein} reported on BM shift in  a series of single crystals with varying level of doping and found $\eg=160$~meV $\pm10\,\%$ at 77~K. Based on an extrapolation, these authors estimated that the 0 K value would be  $175$~meV $\pm10\,\%$. However, it seems that the temperature shift was underestimated. In our data, the shift of the absorption edge between 80 and 10~K is about 35~meV. When taking into account this value, we obtain  $\eg\approx195$~meV at 10~K. In a recent study on thin films\cite{Post2013}, Post \etal\ reported  $\eg=190$~meV for a 99 quintuple layer thick film. The onset of absorption was obtained based on the extrapolation of $\epsilon_2^2$ to zero. If this procedure is used for our data, we obtain $\eg\approx200$~meV, a value closer to the one obtained by Post \etal\cite{Post2013}
From a recent transmission experiment on single crystals, Ohnoutek~\etal\cite{Ohnoutek2016} reported $\eg\approx200$~meV. Notably, a very recent work~\cite{Martinez2017} combining luminescence, transmission and magneto-transport measurements estimated that the gap value is $220\pm5$~meV which agrees within the errorbars with our result.  The values from all these reports fall into an interval 190 -- 220 meV at 10~K and represent a fairly robust estimate for the value of the direct band gap. The differences between the results are presumably due to different ways of how the onset of absorption is defined and how the MB shift is calculated. 

Post \etal\cite{Post2013} noted that these values are significantly smaller than the direct bandgap value of 300~meV determined from some early ARPES data~\cite{Xia2009}. However, in later ARPES data with improved signal to noise ratio, there is a noticeable intensity reaching up to the Dirac point located at about 0.2~eV below the bottom of the conduction band~\cite{Bianchi2010,Chen2010,Bianchi2012,Kuroda2010,ChenPNAS2012,Nechaev2013BiSe}. 
Explicitly, Chen \etal\ concluded that the valence band in \bise\ extends up to the Dirac point from below~\cite{Chen2010}. 
In the view of the more recent ARPES data, it becomes clear that the valence band is a downward oriented paraboloid located at about 0.2~eV below the bottom of the conduction band and thus in accord with the $\approx215$~meV direct gap derived above. The M (or camel-back) structure seen in  ARPES data might be due to surface states or due to a deeper lying structure in the valence bands~\cite{Bianchi2012}. 
The latter scenario is confirmed by recent Shubnikov--de Haas (SdH) measurements~\cite{Piot16}  concluding that the top of the valence band is a downward oriented paraboloid with no signatures of an M  shape of the top of the valence band. 
Note that the M-structure of the valence band seen in some LDA bandstructure calculations~\cite{Zhang10} is absent in the more realistic results of GW calculations~\cite{Yazyev2012,Aguilera2013,Nechaev2013BiSe} that exhibit a paraboloidal shape of the valence band.

\subsection{The absorption edge in \boldmath \bite \unboldmath\ }
\label{BiTeText}

\begin{figure}
	\hspace*{-0.6cm}
	\includegraphics[width=9.8cm]{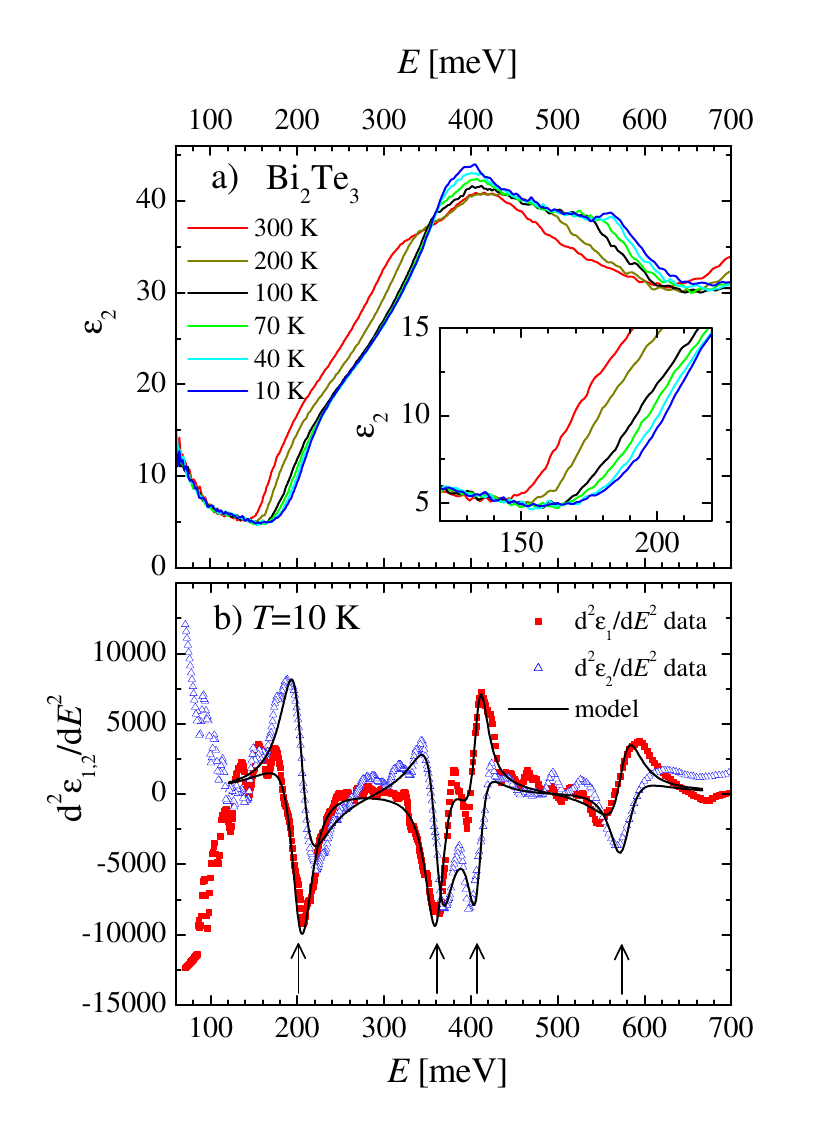}%
	\caption{\label{BiTe} (a) Temperature dependence of the imaginary part of the dielectric function, $\epsilon_2$, of \bite\ sample. The inset shows the spectra on magnified scales. (b) Second derivative of the real (red squares) and imaginary part (blue triangles) of the dielectric function at 10~K together with critical point model spectra (solid lines). The arrows mark the position of four critical points. }
\end{figure}

Figure~\ref{BiTe}(a) displays the temperature dependence of $\epsilon_2$ of our \bite\ thin film. The absorption edge blue-shifts with lower temperature  similarly to \bise. 
In order to obtain the energy of the CPs, we have fitted the second derivative of the dielectric function  with a set of three dimensional CPs of Eq.~(\ref{CP}). Figure~\ref{BiTe}(b) displays the second derivative at $T=10$~K together with a model consisting of four CPs centred at 202, 362, 408 and 575 meV as marked by the arrows. 
At 300~K (not shown), the obtained energy of the lowest CP is 157 meV. 
If the extrapolation of $\epsilon^2_2(E)$ towards zero is used to define $E_{\rm min}$, we obtain the value 190~meV and 152~meV at 10~K and 300~K, respectively. Interestingly, the absorption onset is sharper at 300~K than at 10~K [see the inset of Fig.~\ref{BiTe}(a)], which is the opposite to what is expected from thermal broadening effects. We have checked that this effect is not due to a freezing of residual atmosphere on the sample at low temperature. Obviously below the main interband absorption that sets in near 200~meV at 10~K, an additional weaker interband absorption occurs with an onset at about 170~meV. Albeit weaker than the main interband transition, the magnitude of this absorption is still large (of the order of unity in $\epsilon_ 2$) therefore it most likely corresponds to a direct interband transition. The energy of this onset is the same as found in an earlier report of Sehr and Testardi~\cite{Sehr1962}. 

Chapler~\etal\cite{Chapler2014} reported  the onset of interband absorption in 
Bi$_2$Te$_ 3$ thin films between 140--150~meV at 300~K not too different from our value of 157~meV. Their onset value did not exhibit a significant blue shift with cooling to 10~K. A direct gap was observed in Ref.~\cite{Thomas1992} at 220~meV with a weak onset at 150~meV on relatively highly doped \bite. The difference in the reported values is presumably caused by differences in the doping level and related MB shift and/or by a difference in the definition of the onset energy.
Regardless of these relatively small differences, as noted by Chapler~\etal\cite{Chapler2014}, 
the onset of absorption at 10~K ($\approx200$~meV in our data) is significantly smaller than the 290~meV value of the direct bandgap at the $\Gamma$ point observed in ARPES\cite{Chen2009, Sanchez2014}. 
Note that in our data, the value of $\epsilon_2$ at 290~meV is about 25, which corresponds to a very strong absorption due to direct interband transitions. Obviously, the onset occurs at a different point in the Brillouin zone. Bandstructure calculations\cite{Mishra1997,Zhang10,Nechaev2013,Michiardi2014}  indeed suggest that the direct gap should be close to the Z--F line in the Brillouin zone, where the interband transition energy is significantly smaller than that at the $\Gamma$ point.   The bandgap values have been treated theoretically in detail in Ref.~\cite{Nechaev2013} using the GW calculations. Direct interband transitions are predicted  close to the Z--F line with an onset at 168~meV in a reasonably good agreement with our experimental value. In the calculations, it appears that their location is not exactly on the Z--F line but close to about 1.3Z, see Fig.~3(d) in Ref.~\cite{Nechaev2013}. An indirect fundamental band gap is predicted at a slightly lower energy of 156~meV.

ARPES results suggest that this interpretation is likely. The conduction band in \bite\ has a strong hexagonal warping\cite{Chen2009, Sanchez2014}. The ARPES intensity related to the conduction band is centred at the $\Gamma$ point, however, it has a snowflake-like  shape with protrusions extending into the $\Gamma$--M direction. This direction of the projected surface Brillouin zone measured by ARPES involves the Z--F line of the bulk Brillouin zone. 
Since the valence band has an M shape, its energy increases from the $\Gamma$ towards the M point and reaches a maximum at 
$k\approx0.13$~\AA$^{-1}$ where the difference between valence and conduction band seems to be about 0.2~eV, a value compatible with our optical spectroscopy data. 

According to the GW calculations~\cite{Nechaev2013,Michiardi2014}, the structure of the calculated valence and conduction band along the Z--F line is rather complicated; it seems to have several extrema. It is possible that with changing temperature, the position of the lowest direct bandgap changes and at low temperature, an interband transition with a smaller joint density of states compared to the main one sets in and gives rise to the observed onset at 170~meV. 

We estimate that the BM shift in \bite\ is much smaller than in \bise. The main reason is that, as seen in the ARPES data\cite{Chen2009, Sanchez2014},  the electrons populate first the conduction band states located at the $\Gamma$ point or along the $\Gamma$--Z line~\cite{Michiardi2014} that lie at a smaller energy than those along the Z--F line. As a consequence, the minimum excitation energy is not increased for weakly doped samples. The BM shift likely occurs only for strongly doped samples where the Fermi level enters the conduction bands at the Z--F line. Secondly, the very steep increase of $\epsilon_2$ above the bandgap edge is caused by a quite large joint density of states due to fairly flat bands where only a relatively small BM shift can be expected. 

As a summary for \bite, at~10~K we observed a strong direct interband transition at about 200~meV accompanied with an onset of a weaker direct interband transition at about 170~meV. As suggested by bandstructure calculations, the direct transition occurs near the Z--F line in the Brillouin zone. Since along the $\Gamma$--Z line the energy of the conduction band lies at lower energy than along the Z--F line as suggested by the bandstructure calculations and ARPES~\cite{Nechaev2013,Michiardi2014}, there should be an indirect bandgap with an energy even lower than 170~meV, i.e., \bite\ is an indirect semiconductor. However, these indirect transitions are masked in our data by the much stronger response of free charge carriers that yields of about 5 for $\epsilon_2$ below  $\eg$.  Since typically an indirect gap will not contribute to $\epsilon_2$ with a value larger than about 0.1,  consequently for our thin film samples we cannot observe its contribution to the absorption.

\subsection{The absorption edge in 
	\boldmath
\bitese\ 
	\unboldmath}
\label{BiTeSeText}

\begin{figure}
	\hspace*{-0.8cm}
	\includegraphics[width=9cm]{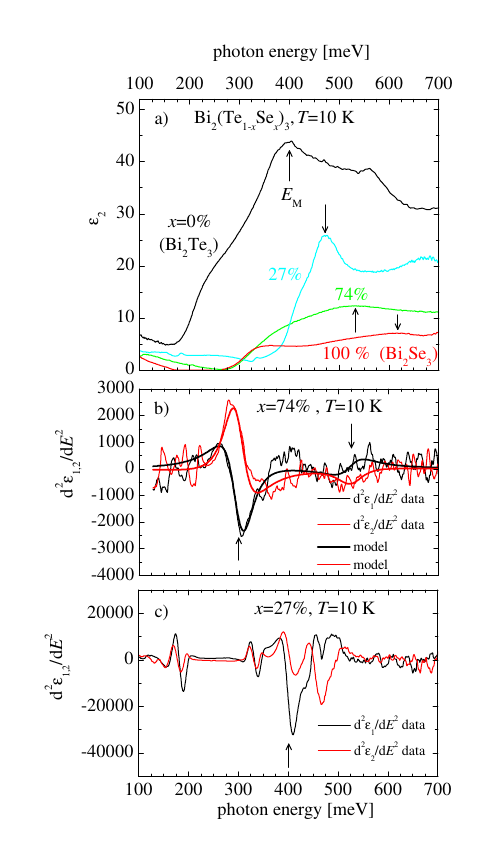}%
	\hspace*{-0.8cm}
	\caption{\label{BiTeSeA} a) The imaginary part of the dielectric function of  our \bitese\ samples at 10~K. The arrows mark the energy energy $E_{\rm M}$ of the maximum of $\epsilon_2$. 
	b)	The second derivative of the real (black thin line) and imaginary part (red thin line) of the dielectric function at 10~K of the $x=74\%$ sample together with the critical point model spectra (thick lines). The arrows mark the position of two critical points. 	
	c) The second derivative of the real (black line) and imaginary part (red line) of the dielectric function at 10~K of the $x=27\%$ sample. The arrow marks the position of the main critical point. The structures near 180 and 330~meV are interference artefacts as discussed in the text.
		}
\end{figure}

Figure~\ref{BiTeSeA}(a) shows $\epsilon_2$ spectra at 10~K of our alloy  $x=27\%$ and 74\% samples. For comparison, the spectra of \bise\ and \bite\ are shown as well.  The spectrum of the $x=27\%$ sample is qualitatively similar to that of \bite, i.e., it exhibits a very steep increase above the onset at 390~meV, reaching high values of $\epsilon_2$ of about 25. 
On the contrary, the spectrum of the  $x=74\%$ sample is qualitatively  similar to that of \bise, i.e., it displays a relatively gradual increase of absorption above the edge up to the values of $\epsilon_2\approx10$. This suggests that the absorption edge occurs at  the same point in the Brillouin zone, i.e, the $\Gamma$ point for the $x=74\%$ sample whereas for the $x=27\%$ sample it is rather at a point close to the Z--F line of the Brillouin zone like as for the \bite. 

We have analysed the spectra in a similar way as described above by fitting the second derivative of $\hat{\epsilon}(E)$ with the CP model of Eq.~(\ref{CP}). 
The second derivative of $\hat{\epsilon}(E)$ for the $x=74\%$ sample is shown in Fig.~\ref{BiTeSeA}(b) together with the model consisting of two CP contributions marked by the arrows. Similarly to \bise, the strong CP centred at 300~meV is modelled with a two dimensional critical point and the weak one at 525~meV with a three dimensional CP. Using the same  formulas as for the \bise\ sample, the BM shift amounts to 33~meV and the bandgap value is $270\pm10$~meV. 

\begin{figure}
	\hspace*{-0.8cm}
	\includegraphics[width=10cm]{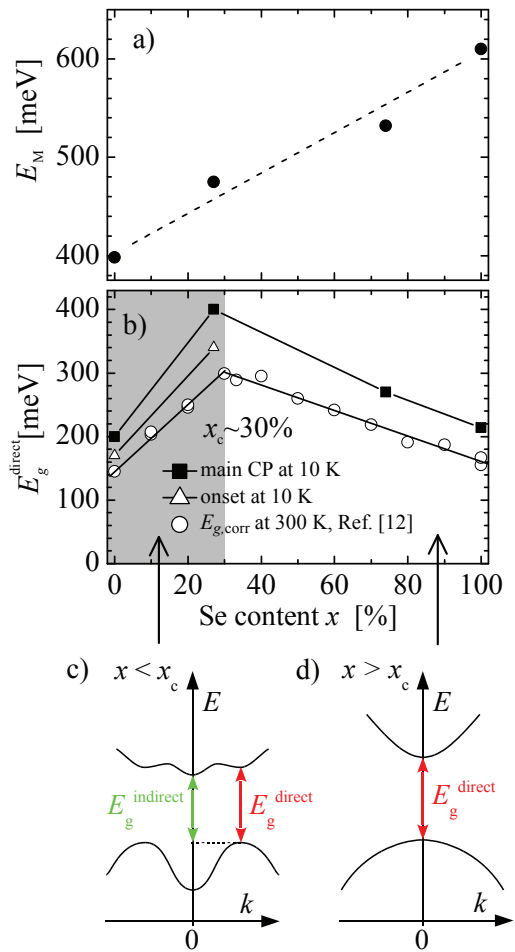}%
	\caption{\label{BiTeSeB} a)  The energy $E_{\rm M}$ of the maximum of $\epsilon_2$ of \bitese\ marked by the arrows in Fig.~\ref{BiTeSeA}. The dashed line is a guide to the eye. b) Extracted values of the onset of direct interband transitions, $\eg$, at 10~K defined as the centre position of the main lowest critical point (squares). The values for $x=76\%$ and $x=100\%$ were corrected for the Burstein-Moss shift as described in the text. The triangles represent the onset of a weaker direct interband absorption below the main critical point. The circles are previously reported~\cite{Greenaway1965} absorption edge energies at 300~K corrected for the Burstein-Moss effect. c) and d) schematically represent the position of the direct interband transitions (red arrows) in the band structure for Se content $x<x_c$ (c) and $x>x_c$ (d), respectively, where $x_c\approx30\%$. The green arrow in (c) depicts the indirect fundamental gap (see Refs.~\cite{Nechaev2013,Michiardi2014}). }
\end{figure}

The second derivative of $\hat{\epsilon}(E)$ of the $x=27\%$ sample is shown in 
Fig.~\ref{BiTeSeA}(c). Unfortunately, here the spectra, similarly to our \bise\  sample, exhibit near 180~meV and 330~meV artefacts that correspond to interference fringes [see Fig.~\ref{elli}(b)]. These artefacts prevent us from  fitting the data with the CP model;  nevertheless, it is clear that the strongest feature centred at about 400~meV (marked by the arrow) corresponds to the strongest CP related to the onset of the direct interband transitions. Similarly to \bite, below this CP a weaker absorption occurs with the onset at about 340~meV, see Fig.~\ref{BiTeSeA}(a). The plasma frequency of this sample is even smaller than that of the \bite\ sample (see Tab.~\ref{Table}) and thus, similarly to \bite, we estimate that the BM shift is either absent or negligibly small. 
Both the CP energy of 400~meV and the onset at 340~meV are significantly larger than the bandgap value of $290$~meV reported\cite{Reijnders2014} for a sample with similar Se content $x=33\%$. However, the energy of the maximum of $\epsilon_2$ denoted here as $E_{\rm M}$ [see the arrows in Fig.~\ref{BiTeSeA}(a)] occurs at different energy, namely at 520~meV at 10~K in case of Ref.\cite{Reijnders2014} whereas it is at 475 meV in our data. Since the energy of this maximum strongly increases with increasing $x$ [see Fig. ~\ref{BiTeSeB}(a)], it is likely that the composition of the two samples is different, the one of Ref.\cite{Reijnders2014} having significantly larger value of $x$ than our sample. We attribute the maximum $E_{\rm M}$ to a CP related to the Z--F line whose energy monotonically increases with  Se content.

Figure~\ref{BiTeSeB}(b) displays the extracted values of the onset of direct interband transitions, $\eg$, at 10~K
of all our samples. The squares correspond to the main CP energies. For the $x=74\%$ and 100\% samples these were corrected for the BM shifts, while for the other two samples, we expect that the BM shift is absent or small. The triangles correspond to a weaker absorption  observed in the spectra of the $x=0\%$ and 27\%  samples that are likely due to a weaker direct interband transition as discussed above. In general, the dependence of the bandgap value on selenium content has a roof-like shape with a  maximum for the $x=27\%$ sample. This shape is very similar to the one reported for bandgap values at 300~K corrected for the Burstein-Moss effect in the early work of Greenaway and Harbecke\cite{Greenaway1965}, which are for comparison added to Fig.~\ref{BiTeSeB}(b) as circles. 
The latter dependence exhibits a pronounced maximum of the direct gap near $x=30\%$ content; our values are of course larger since they were obtained at 10~K. Similarly to their conclusion, we suggest that the direct interband transitions on the left and right side of the maximum corresponds to different points in the Brillouin zone. The position of the onset of direct interband transitions in the band structure for the two cases is schematically represented in Figs.~\ref{BiTeSeB}(c)  and \ref{BiTeSeB}(d) by red arrows. Figure~\ref{BiTeSeB}(d) depicts direct bandgap in the centre of the Brillouine zone between parabolic-like bands that occurs on the \bise\ side of the \bitese\ alloy. On the \bite\ side shown in Fig.~\ref{BiTeSeB}(c), the direct bandgap occurs off the Brillouine zone centre near the Z--F line,  between the wing of the M-shaped valence band and a conduction band~\cite{Nechaev2013,Michiardi2014}. The green arrow depicts the indirect fundamental gap which, however, has orders of magnitude lower absorption and therefore is masked in our data by the much stronger free charge carriers response. 

\section{Conclusion}
We have examined the optical response of \bitese\ thin films which reveal the direct interband transitions. In \bise, after the correction for the Burstein-Moss effect, we find a bandgap value
of $215\pm10$~meV at 10~K in good agreement with the values from 190 to 220 meV reported in the literature\cite{Kohler74Burnstein,Post2013,Ohnoutek2016,Martinez2017}.  Our result supports the conclusion  of ARPES~\cite{Chen2010}, GW calculations~\cite{Yazyev2012,Aguilera2013} and SdH measurements~\cite{Piot16} that the direct bandgap of \bise\ is located at the $\Gamma$ point in the Brillouin zone, the valence band reaches up to the Dirac point and has a shape of a downward oriented paraboloid. 

In \bite, we observe a strong interband absorption that sets in at $\approx200$~meV at 10~K and a weaker albeit still direct interband transition with an onset at about 170~meV. These values support GW band structure calculations which proposed that the direct interband transition occurs near the Z--F line. 
In the ternary \bitese\ alloys the energy of the onset of direct interband transition goes through a maximum for a Se content of about 30\%. This behaviour indicates that there is a crossover of the position of the direct interband transitions between the two different points in the Brillouin zone at this Se content. 

%\appendix
%\section{}

\begin{acknowledgments}
We acknowledge helpful discussions with I. Aguilera. This work was financially supported from European Regional Development Fund Project CEITEC Nano+ (No. CZ.021.01/0.0/0.0/16\_013/0001728),
CEITEC Nano Research Infrastructure (ID LM2015041, MEYS CR, 2016–2019), CEITEC Brno University of Technology and by EC via TWINFUSYON project (No. 692034).
\end{acknowledgments}

\bibliography{bibliography}

\end{document}